# Crossover from 2D metal to 3D Dirac semimetal in metallic PtTe$_2$ films with local Rashba effect

Ke Deng,[1,*] Mingzhe Yan,[1,*] Chu-Ping Yu,[2] Jiaheng Li,[1] Xue Zhou,[1] Kenan Zhang,[1] Yuxin Zhao,[1] Koji Miyamoto,[3] Taichi Okuda,[3] Wenhui Duan,[1,4] Yang Wu,[5] Xiaoyan Zhong,[2] and Shuyun Zhou[1,4,†]

[1]*State Key Laboratory of Low Dimensional Quantum Physics and Department of Physics, Tsinghua University, Beijing 100084, China*
[2]*National Center for Electron Microscopy in Beijing, Key Laboratory of Advanced Materials (MOE), State Key Laboratory of New Ceramics and Fine Processing, School of Materials Science and Engineering, Tsinghua University, Beijing 100084, China*
[3]*Hiroshima Synchrotron Radiation Center, Hiroshima University, Higashi-Hiroshima 739-0046, Japan*
[4]*Collaborative Innovation Center of Quantum Matter, Beijing 100084, China*
[5]*Department of Physics and Tsinghua-Foxconn Nanotechnology Research Center, Tsinghua University, Beijing 100084, China*

*These authors contributed equally to this work.
†Correspondence should be sent to syzhou@mail.tsinghua.edu.cn

**Abstract:**

PtTe$_2$ and PtSe$_2$ with trigonal structure have attracted extensive research interests since the discovery of type-II Dirac fermions in the bulk crystals. The evolution of the electronic structure from bulk 3D topological semimetal to 2D atomic thin films is an important scientific question. While a transition from 3D type-II Dirac semimetal in the bulk to 2D semiconductor in monolayer (ML) film has been reported for PtSe$_2$, so far the evolution of electronic structure of atomically thin PtTe$_2$ films still remains unexplored. Here we report a systematic angle-resolved photoemission spectroscopy (ARPES) study of the electronic structure of high quality PtTe$_2$ films grown by molecular beam epitaxy with thickness from 2 ML to 6 ML. ARPES measurements show that PtTe$_2$ films still remain metallic even down to 2 ML thickness, which is in sharp contrast to the semiconducting property of few layer PtSe$_2$ film. Moreover, a transition from 2D metal to 3D type-II Dirac semimetal occurs at film thickness of 4–6 ML. In addition, Spin-ARPES measurements reveal helical spin textures induced by local Rashba effect in the bulk PtTe$_2$ crystal, suggesting that similar hidden spin is also expected in few monolayer PtTe$_2$ films. Our work reveals the transition from 2D metal to 3D topological semimetal and provides new opportunities for investigating metallic 2D films with local Rashba effect.

# 1. Introduction

Transition metal dichalcogenides (TMDCs) are fascinating layered materials with rich physics and potential applications [1-3]. Among the large family of compounds, group-10 TMDCs (PtS$_2$, PtTe$_2$ and PtSe$_2$) with trigonal (1T) structure have distinctive properties, yet they remain least investigated until recent years [4-8]. In contrast to semiconducting group-6 TMDCs (e.g. MoS$_2$) which usually forms thermodynamically stable hexagonal (2H) structure, platinum ditelluride (PtTe$_2$) crystalizes in trigonal structure and early band structure calculation suggests that the bulk crystal is a metal [9]. Recently, bulk PtTe$_2$ crystal has been reported to host nontrivial topological type-II Dirac fermions, and massless Dirac fermions are found to emerge at the topologically protected touching points of electron and hole pockets [6]. The strongly tilted Dirac cone along the out-of-plane momentum direction breaks the Lorentz invariance and therefore provides the counterpart beyond standard model of physics [10]. The isostructural material PtSe$_2$ is also a type-II Dirac semimetal [11,12], and shows metal-semiconductor transition with decreasing thickness [4, 13, 14]. In addition, monolayer PtSe$_2$ shows interesting helical spin texture [8] induced by local Rashba effect (R-2) [15] despite the fact that the monolayer film itself is centrosymmetric. Such hidden spin texture induced by local Rashba effect has been predicted to provide an important platform for realizing novel topological superconductivity with odd parity if s-wave superconducting pairing can be added [16-18]. Theoretically it has been proposed that carrier doping on PtSe$_2$ can tune the electron-phonon interaction and enhance the strength of superconductivity [19]. However, considering that PtSe$_2$ is a semiconductor with a large gap size of ≈1.2 eV [4], it is difficult to tune the Fermi energy in such a large range to make it a superconductor. Finding a similar centrosymmetric film with local Rashba effect yet with metallic property can provide a better opportunity for realizing topological superconductivity, and PtTe$_2$ is a good candidate considering that tellurides are usually more metallic than selenides.

Growth of PtTe$_2$ thin films with thickness down to a few nanometers has been reported by chemical vapor deposition (CVD) recently [20-22]. However, so far there is no report of thinner (sub-nanometer) films and the experimental electronic structure of PtTe$_2$ films still remains to be explored. In this work, we report a systematic angle-resolved photoemission spectroscopy (ARPES) study of the electronic structure of PtTe$_2$ films grown by molecular beam epitaxy (MBE) with controlled film thickness from 2 monolayers (ML) to 6 ML. ARPES measurements show that the PtTe$_2$ films remain metallic even down to the thinnest 2 ML, and there is a transition from 2D metallic film to 3D type-II Dirac semimetal at 4–6 ML. Spin-ARPES measurements reveal the spin texture induced by local Rashba effect in bulk PtTe$_2$ crystal. Considering that the local Rashba effect has been observed in bulk PtTe$_2$, PtSe$_2$ and monolayer PtSe$_2$, which all have similar crystal symmetry to PtTe$_2$ films, we expect PtTe$_2$ films to exhibit similar local Rashba effect. Therefore, atomically thin PtTe$_2$ film is an interesting metallic film with local Rashba effect, which is compatible with requirements for the new mechanism of topological superconductivity [16].

## 2. Method

Atomically thin PtTe$_2$ films with different thickness from 2 ML to 6 ML were grown on bilayer graphene/6H-SiC (0001) in the ultra-high vacuum (UHV) system with a base pressure of $2 \times 10^{-10}$ torr. The graphene was grown by flash annealing the 6H-SiC substrate at 1,350 °C [23]. The growth process is monitored by high energy electron diffraction (RHEED). After 60 cycles of flash annealing process, sharp diffraction pattern from graphene is observed in low energy electron diffraction (LEED) (Fig. 1c) and RHEED (Fig. 1e). High purity Pt (Alfa Aesar, 99.95%) and Te (Alfa Aesar, 99.999%) were then evaporated onto the substrate at 300 °C with the flux ratio of 20:1 to ensure a Te rich environment. After the film growth, in situ annealing process was employed to improve the film quality, and the Te flux was maintained during the annealing process to avoid Te vacancies in the sample. The as-grown PtTe$_2$ films with different thickness were then transferred to angle-resolved photoemission spectroscopy (ARPES) chamber for in situ ARPES measurements, or capped with Te protection layer before taking out for TEM measurements. ARPES measurements were performed at 80 K using helium lamp (21.2 eV) source. Spin-ARPES measurements of PtTe$_2$ bulk crystal were performed at ESPRESSO endstation of Hiroshima Synchrotron Radiation Center at temperature of 20 K using Helium lamp.

First-principle calculations were carried out by the Vienna ab initio Simulation Package (VASP) [24] in the framework of density functional theory, with the projector augmented wave (PAW) methods and the Perdew-Burke-Ernzerhof-type generalized gradient approximation (GGA). The kinetic energy cutoff used in the whole calculations is fixed at 400 eV, and the DFT-D3 method is included to correct van der Waals interaction of layered PtTe$_2$. The spin-orbit coupling effect is included self-consistently. The $k$-point mesh grid of $12 \times 12 \times 6$ is taken for bulk calculations, and $12 \times 12 \times 1$ for multi-layer thin films. The lattice constants are fixed at experimental data, and all the atomic positions are fully relaxed with the force criteria of 0.01 eV/Å.

## 3. Results and discussion

PtTe$_2$ crystallizes in trigonal structure (see top and side views in Fig. 1a), corresponding to CdI$_2$-type structure with $P\bar{3}m1$ space group. Sharp diffraction pattern from the PtTe$_2$ film is observed both in LEED and RHEED (Fig. 1d,f) after PtTe$_2$ film growth, indicating the high quality of the film. The crystal structure of PtTe$_2$ film is revealed by transmission electron microscopy (TEM) measurements in Fig. 1g. The high quality MBE PtTe$_2$ films with controllable thickness down to 2 ML provide a unique opportunity for a systematic study of the electronic structure as a function of film thickness.

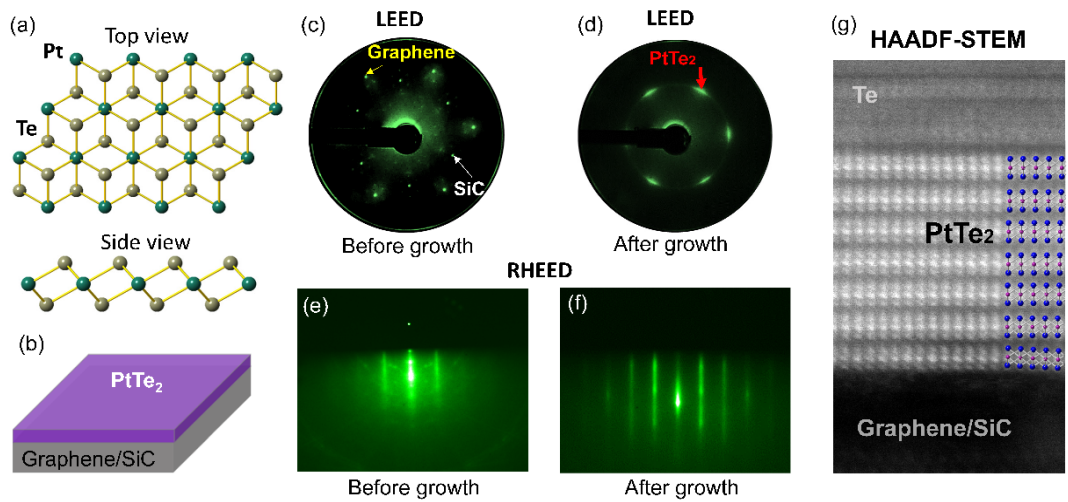

**Fig.1.** Crystal structure of 1T-PtTe$_2$ film. (a) Top and side views of PtTe$_2$. Each monolayer is defined as one PtTe$_2$ sandwiched block. (b) An illustration of PtTe$_2$ grown on bilayer graphene/6H-SiC (0001). (c, d) LEED patterns of sample before (c) and after (d) growth respectively. (e, f) RHEED patterns taken before (e) and after (f) PtTe$_2$ film growth respectively. (g) HAADF-STEM image of a sample cross section. Te atoms were capped over the sample surface to protect it from degradation.

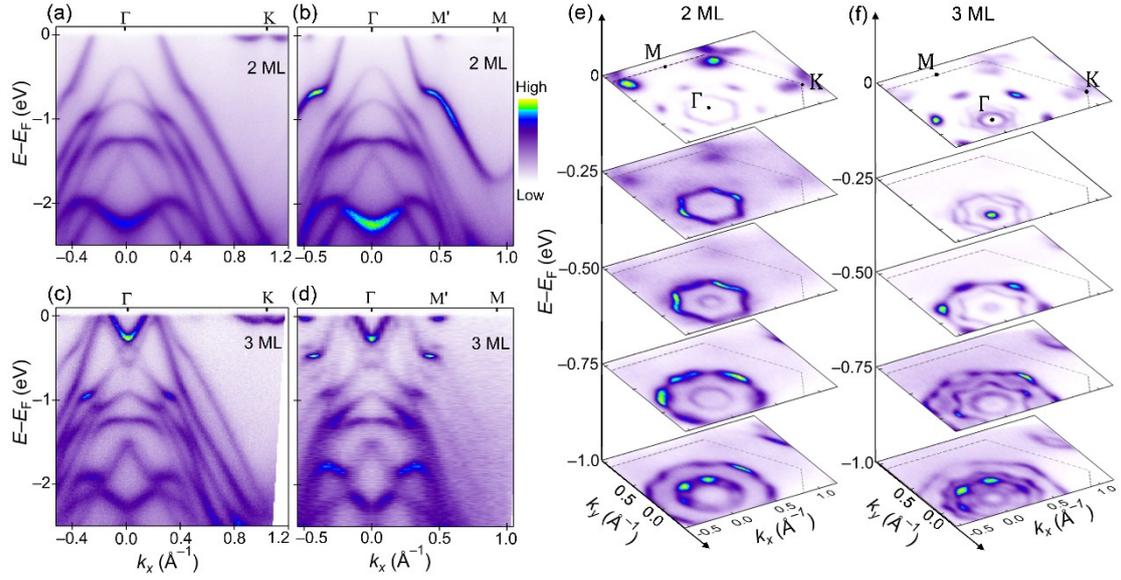

**Fig. 2.** The electronic structure of 2 ML and 3 ML PtTe$_2$ films (a, b) ARPES spectra of 2 ML PtTe$_2$ film along the $\Gamma - K$ and $\Gamma - M$ directions, respectively. (c, d) ARPES spectra of 3 ML PtTe$_2$ film along the $\Gamma - K$ and $\Gamma - M$ directions, respectively. (e, f) Intensity maps measured at energies from $E_F$ to –1.0 eV.

Fig. 2 shows the electronic structures of 2 ML and 3 ML PtTe$_2$ films measured by ARPES. Fig. 2a–d shows the dispersions for 2 ML and 3 ML films along two high symmetry directions, $\Gamma - K$ and $\Gamma - M$ direction respectively. For 2 ML film, the dispersion shows a hole pocket centered at the $\Gamma$ poiont near the Fermi energy ($E_F$), and two electron pockets near the K point. Along the $\Gamma - M$ direction, another electron pocket is observed at the mid-point (labeled as M′) between the $\Gamma$ and M points. Such pocket is observed as gapped Dirac cone in bulk PtTe$_2$ [6]. The electronic structure shows that the 2 ML film is metallic, which is in sharp contrast to PtSe$_2$ films [13]. For 3 ML film, the pocket at the $\Gamma$ point splits into two, and there are more bands at high binding energies as well. In addition, there are two electron pockets centered at the $\Gamma$ point. Fig. 2e, f shows the intensity maps at selected energies from $E_F$ to –1 eV measured for both 2 ML and 3 ML films. The electron pocket at the $\Gamma$ point shows a hexagonal shape at $E_F$ and it expands in size with binding energy in 2 ML sample. Below –0.5 eV, another pocket starts to appear at the $\Gamma$ point and it expands in size with energy too. In addition, there are electron pockets around the K point and the M′ point at $E_F$. For 3 ML film, there are four electron pockets at $E_F$ and they become more obvious at –0.25 eV, and the pockets near the K and M′ points are similar to those in 2 ML film. The ARPES measurements show that unlike its isostructural PtSe$_2$ films that show metal-semiconductor transition with decreasing film thickness [4, 13], PtTe$_2$ films, however, remain metallic even down to 2 ML. First principle calculations show that 1 ML PtTe$_2$ is semiconductor [25-27], which still awaits further experimental verification.

To reveal the evolution from 2D metal film to 3D type-II Dirac semimetal, we show in Fig. 3 a systematical study of the band structure of PtTe$_2$ films from 2 ML to 6 ML and the bulk crystal along the Γ − K direction. There are two interesting conical dispersions and we can follow their evolution with sample thickness. First of all, from 2 ML to 3 ML, two V-shaped pockets (marked by red arrow in Fig. 3b) emerge at $E_F$ at the Γ point. The V-shaped pockets move down in energies and eventually touch the hole-like pocket at higher binding energy, resulting in three-dimensional type-II Dirac fermions in the bulk (Fig. 3e) [6]. We note that the 4 ML and 6 ML films show similar dispersion to the bulk crystal, and this suggests that they effectively have the bulk property of the topological semimetal. The thickness dependent ARPES data allow to follow the evolution of the valence band and reveal the transition from a 2D metal to a 3D topological semimetal at thickness of 4–6 ML. The other conical dispersion is at higher energy from –2 to –3 eV (grey arrows). For 2 ML film, there is a gap between the rounded M-shaped and W-shaped dispersions between –2 and –3 eV. These two bands become closer in 3 ML film and they merge to form a Dirac cone in 4 ML and 6 ML films. This Dirac cone is the topological surface state, similar to that reported in PtSe$_2$ [6] and PdTe$_2$ [28]. First principles calculation (dotted curves in Fig. 3f–j) shows dispersions in agreement with ARPES results, supporting the transition from 2D metallic film to 3D type-II Dirac semimetal at 4–6 ML.

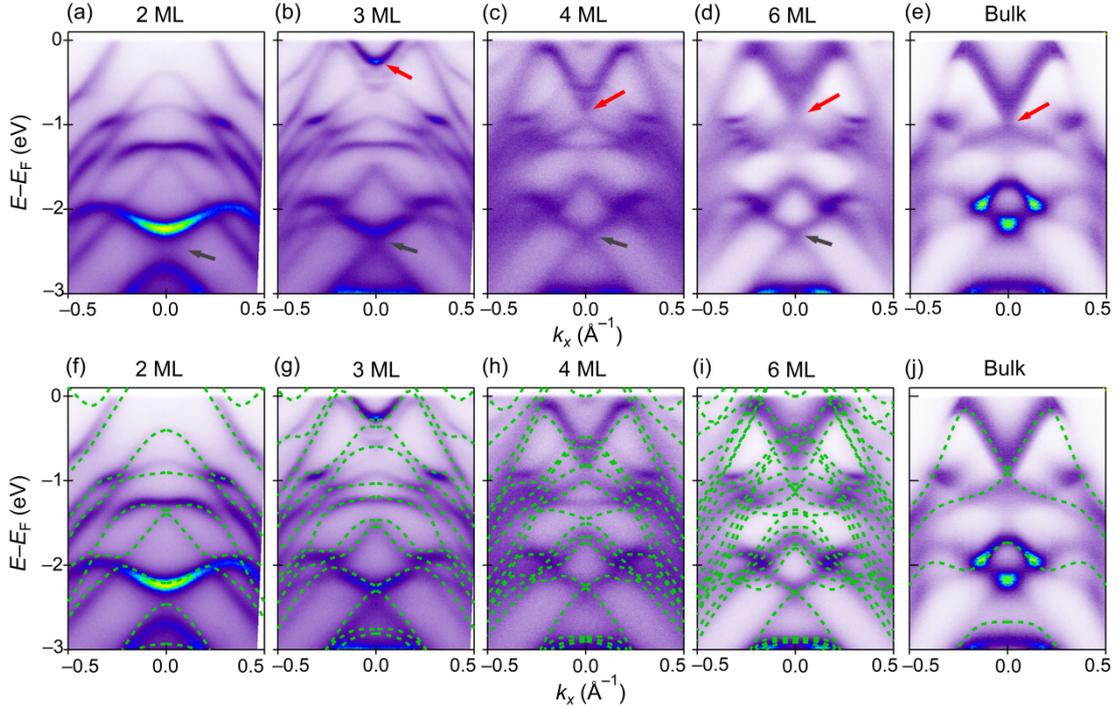

**Fig. 3.** Evolution of the electronic structure of PtTe$_2$ films from 2D metal to 3D type-II Dirac semimetal. (a–e) Measured ARPES spectra of PtTe$_2$ films with different thickness from 2 ML to 6 ML and the bulk crystal along the Γ − K direction. (f–j) Comparison of measured ARPES dispersions with calculated band structures (green dotted lines) along the Γ − K direction.

Considering that PtTe$_2$ and PtSe$_2$ have similar crystal structure, and that helical spin texture is observed both in monolayer PtSe$_2$ film [8] and bulk PtSe$_2$ [13] due to the local Rashba effect, we would expect similar spin physics also applies to bulk PtTe$_2$ and few layers of PtTe$_2$ films.

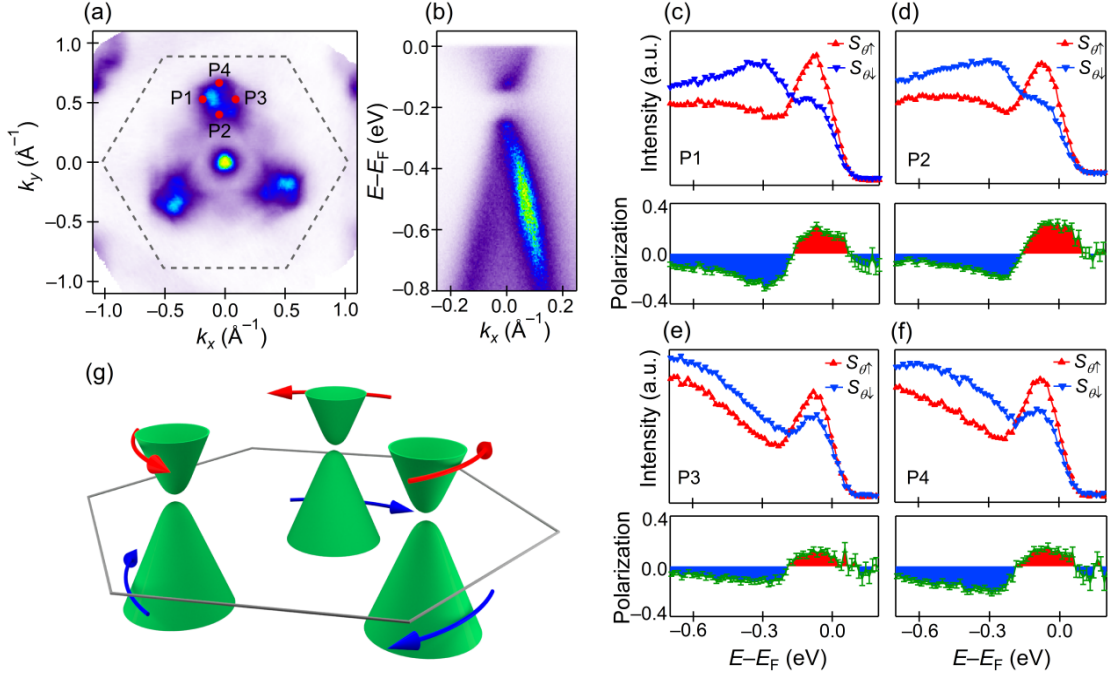

**Fig.4.** Helical spin texture for Dirac-like dispersion in the M′ point of single crystal PtTe$_2$. (a) Intensity map measured at –0.6 eV. Red dots mark the measurement position P1–P4 for EDCs shown in (c–f). (b) Measured dispersion of the gapped Dirac-like dispersion at photon energy of 21.2 eV. (c–f) Spin-resolved EDCs for the in-plane tangential direction from P1–P4. (g) A schematic plot of the spin texture of the gapped Dirac-like dispersion.

In Fig. 4, we reveal the helical spin texture for bulk PtTe$_2$ crystal. Fig. 4a shows the intensity map of PtTe$_2$ at –0.6 eV and a pocket is observed at the M′ point. The dispersion of this pocket shows a gapped Dirac-like dispersion (Fig. 4b). Fig. 4c–f shows the spin contrast of energy distribution curves (EDCs) along the in-plane tangential direction at four selected momentum points on this pocket (P1 to P4). A large spin contrast is observed along the tangential direction ($\theta$) for all EDCs. While the lower part of Dirac-like dispersion shows spin-down at all points measured (P1–P4), the upper cone is spin-up with strong polarization. Fig. 4g shows an illustrative figure of the observed spin polarization. Here all these points on the same Dirac-like dispersion at the M′ point show spin polarization along the tangential direction with respect to Γ point. Considering the three-fold rotational symmetry of the crystal, the spin polarization of these three cones at the three equivalents M′ points forms a helical spin texture with respect to the Γ point. Such in-plane tangential helical spin texture can be explained by the local dipole field between the Pt and Te atoms, similar to the case of PtSe$_2$ [8, 13]. Since the crystal symmetry of the bulk PtTe$_2$ is the same as atomically thin PtTe$_2$ films, the PtTe$_2$ films from 2 ML to 6 ML are also expected to have similar local Rashba effect and similar spin texture. Similar hidden spin polarization has also been predicted in other 1T films [29], which still awaits further experimental verification. Here we show that atomically thin $PtTe_2$ is metallic with local Rashba effect, and can be tuned to 3D Dirac fermions by increasing the film thickness.

## 4. Conclusions

In summary, we have successfully grown atomically thin $PtTe_2$ films with controlled thickness and revealed the evolution of the electronic structure with film thickness. Our ARPES results show that even the 2 ML $PtTe_2$ film is metallic, which is in contrast to semiconducting $PtSe_2$. Moreover, a systematic study shows that the $PtTe_2$ film evolves from a 2D metallic film to a 3D Dirac semimetal at 4–6 ML thickness. In addition, we report helical spin texture induced by the local Rashba effect in bulk $PtTe_2$, which also holds in atomically thin metallic $PtTe_2$ films (including 2 ML, 4 ML to 6 ML films) due to the same crystal structure. Metallic $PtTe_2$ thin films with local Rashba effect, can serve as a new platform to investigate topological superconductivity. Our work reveals the novel physics in $PtTe_2$ and paves the way for further investigation.

## Conflict of interest

The authors declare that they have no conflict of interests.

## Acknowledgments

This work was supported by the National Natural Science Foundation of China (11725418 and 11334006), Ministry of Science and Technology of China (2016YFA0301004, 2016YFA0301001, and 2015CB921001), Science Challenge Project (TZ2016004) and Beijing Advanced Innovation Center for Future Chip (ICFC). Spin-ARPES experiments at Hiroshima Synchrotron Radiation Center have been performed under the proposal number 14-A-15 and 16AG058.

## Author contributions

SZ conceived the research project. MY, XZ and YZ grew and characterized the thin film samples, KZ grew the $PtTe_2$ single crystal with support from YW. MY, XZ, YZ, KZ and KD performed the ARPES measurements and analyzed the data. ZY and XYZ performed the TEM measurements. JL and WD performed first-principle calculations. KD, MY and SZ wrote the manuscript, and all authors commented on the manuscript.